\documentclass[jkps,twoside,twocolumn,epsfig ]{revtex4}
\usepackage{graphicx}

\begin{document}
\title[] {Herd Behaviors in the Stock and Foreign Exchange Markets }
\author{Kyungsik \surname{Kim}}
\email{Kskim@pknu.ac.kr ; fax:+82-51-611-6357}
\author{Seong-Min  \surname{Yoon$^{\dagger}$}}
\author{Yup \surname{Kim$^{\ddagger}$}}
\affiliation{
$^{*} $ Department of Physics, Pukyong National University, Pusan 608-737, Korea \\
$^{\dagger}$Division of Economics, Pukyong National University,\\
 Pusan 608-737, Korea\\
$^{\ddagger}$Department of Physics and Research Institute for Basic Sciences, Kyung-Hee University, Seoul 130-701,
and Asia Pacific Center for Theoretical Physics, Seoul, Korea \\
}
\received{ April 2003}
\begin{abstract}
The herd behaviors of returns for the won-dollar exchange rate and the KOSPI
are analyzed in Korean financial markets.
It is shown that the probability distribution $P(R)$ of price returns $R$ for three values of 
the herding parameter tends to a power-law behavior
$P(R) \simeq R^{-\beta}$ with the exponents $ \beta=2.2$(the won-dollar exchange rate)
and $2.4$(the KOSPI).
The financial crashes are found to occur at $h >2.33$ when the relative increase in the 
probability distribution of exteremely high price returns is observed.
Especially, the distribution of normalized returns shows
a crossover to a Gaussian distribution for the time step
$\Delta t=252$.
Our results will be also compared to the other well-known analyses.\\
\\
PACS: 02.50.-r, 02.50.Ey, 02.50.Wp, 89.90.+n \\

\end{abstract}

\maketitle

\section{Introduction}

There have been considerable interests$[1-3]$ in the microscopic models
for the financial markets.
The major models which are based on the self-organized phenomena
are the herding multiagent model$[4,5]$ and the related percolation models$[6,7]$, the democracy and dictatorship model$[8]$,
self-organized dynamical model$[9]$, the cut and paste model, the fragmentation and
coagulation model$[10]$.
It has been well-known that one of very important microscopic models is the herding model$[11,12]$, in which
some degrees of coordination
among a group of agents share the same information or the same rumor and
make a common decision in order to create and produce the returns.
Recently, a theoretical model$[5]$ for the herding behavior
has been suggested. In this model the probability distribution of returns
shows a power-law behavior for varying values smaller than a critical herding parameter
value, while an relative increase in the probability of large returns
is observed in the region of financial crashes
for the herding parameter larger than the critical value.
In particular, the distribution of normalized returns has
the form of the fat-tailed distributions$[13]$ and
a crossover toward the Gaussian distribution can be shown in financial markets.

On the other hand, the theoretical models and numerical analyses
for the volume of bond futures transacted at Korean futures exchange market
were presented in the previous work$[14]$.
In that work we mainly considered the number of transactions for two
different delivery dates and found
the decay functions for survival probabilities$[15,16]$ in our analyses
of bond futures.
We also studied the tick dynamical behavior of the bond futures price using
the range over standard deviation or the R/S analysis in Korean futures exchange market$[17]$.
The recent work$[18]$ on Norwegian and US stock markets has shown that there exists the
notable persistence caused by long-memory in the time series.
The numerical analyses based on
multifractal Hurst exponent and the height-height correlation
function have also been used mainly for the long-run memory effects. It was
particularly shown that the form of the probability distribution of the normalized return
leads to the Lorentz distribution rather than the Gaussian distribution$[17]$.

The purpose of this paper is to study the dynamical herding behavior 
for the won-dollar exchange rate and the KOSPI(Korean stock price index) in Korean financial markets.
In section $2$ the financial crashes and the distribution of normalized returns
for the tick data of two different delivery dates are analyzed numerically.
The results and conclusions are given in the final section.

\section{Financial Crashes and Simulations}

In our analyses, we introduce the won-dollar exchange rate and the KOSPI
in Korean financial markets.
In this paper, we only consider two delivery dates: 
The tick data for the won-dollar exchange rate were taken from April $1981$ to December $2002$,
while we used the tick data of the KOSPI transacted for $23$ years from April $1981$.
We show the time series of the won-dollar exchange rate $P(t)$ in Fig.$1$, and 
the price return $ R_{\tau} (t)$ is defined as
\begin{equation}
R_{\tau } (t ) =  \ln \frac{P(t + \tau )} { P(t )} ,
\label{eq:a1}
\end{equation}
where $\tau$ denotes the time interval. 
Our average time $\tau$ between ticks is about one day in two types of our tick data,
as shown in Fig.$1$ and $2$.
\begin{figure}[]
\includegraphics[width=8.5cm]{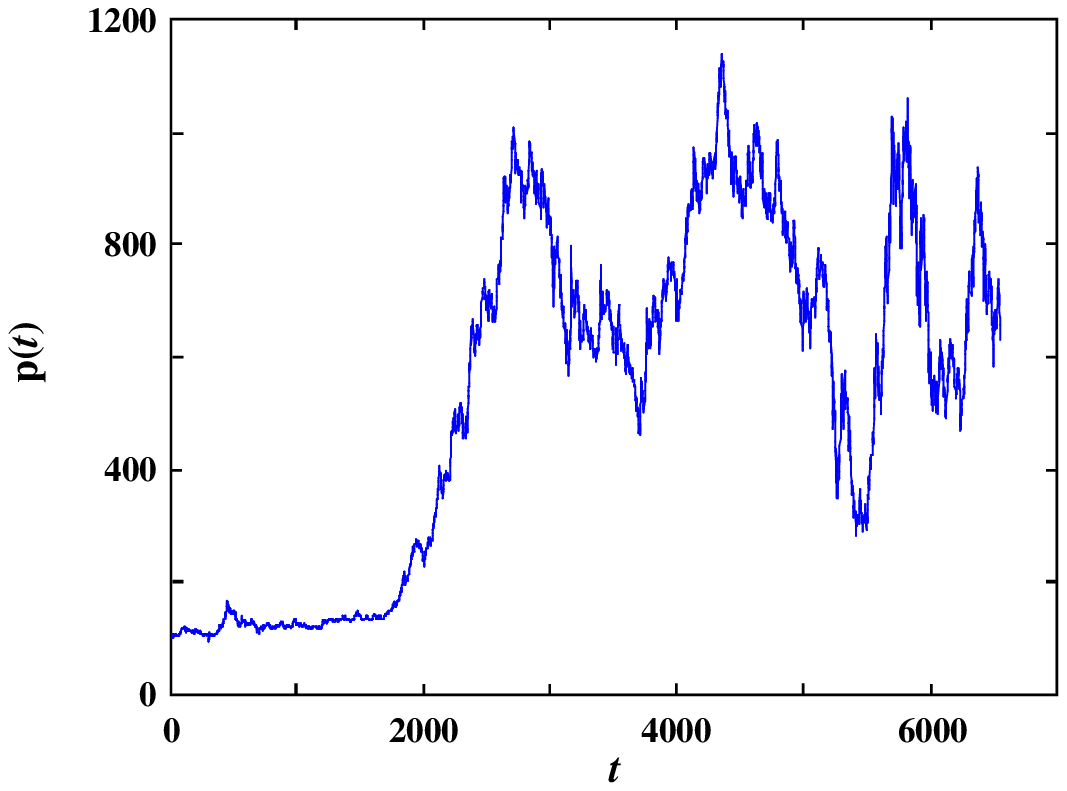}
\caption[0]{Time series of the price $P(t)$ for the KOSPI,
where we used the tick data of the KOSPI transacted for 23 years from April $1981$.}
\end{figure}
\begin{figure}[]
\includegraphics[width=7.5cm]{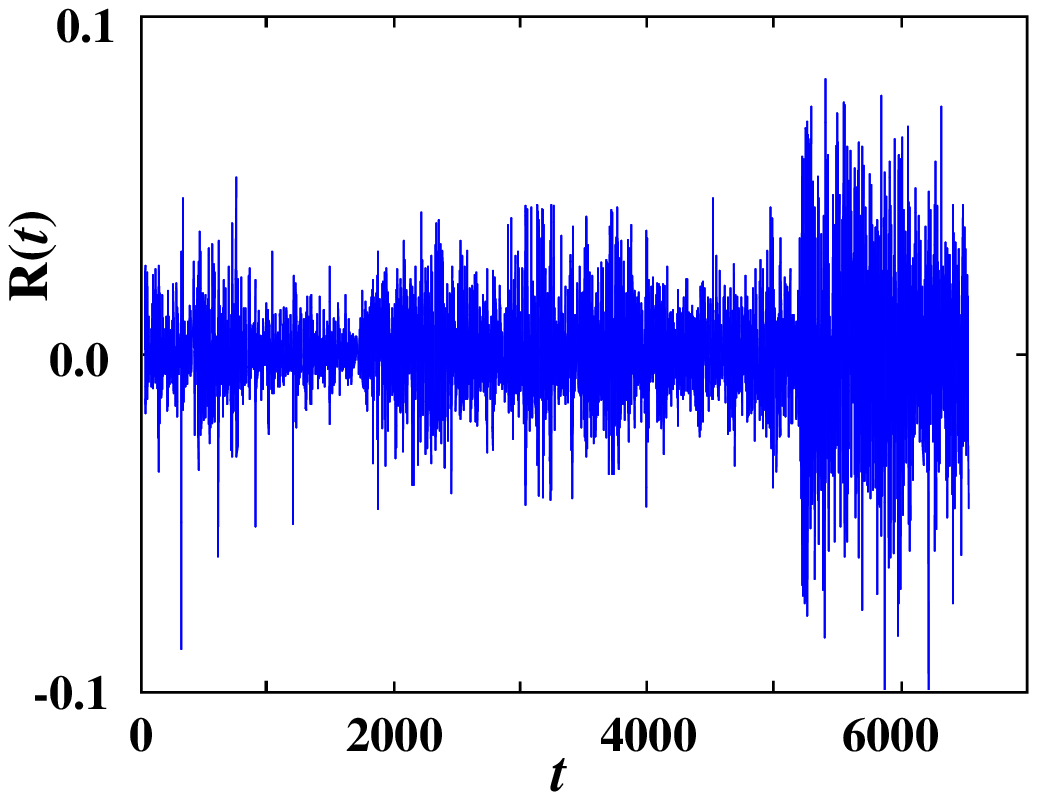}
\caption[0]{ Plot of the price return $R_{\tau=1} (t ) = $ ln$[P(t +1 )/P(t )]$ for the KOSPI.}
\end{figure}
From now on, in order to describe the averaged distribution of
cluster, we let consider three return states composed by $N$
agents, i.e., the continuous tick data of the won-dollar exchange rate and the KOSPI. 
From the states of agent $l$ composed of the three
states $\psi_l = \lbrace -1, 0, 1 \rbrace $, the state of clusters is given by 
\begin{equation}
s(t ) =   \sum_{l=1}^{N} \psi_l ,
\label{eq:b1}
\end{equation}
where $\psi_l ={0} $ is the waiting state that occurs no transactions or gets no return
and $\psi_l ={1}$ ($\psi_l ={-1}$) is the selling (buying) states,
i.e., the active states of the transaction.
Assuming that it belongs to the same cluster between a group of agents sharing the
same information and making a common decision,
the active states of transaction can be represented by vertices
in a network having links of time series.
\begin{figure}[b]
\includegraphics[width=8.5cm]{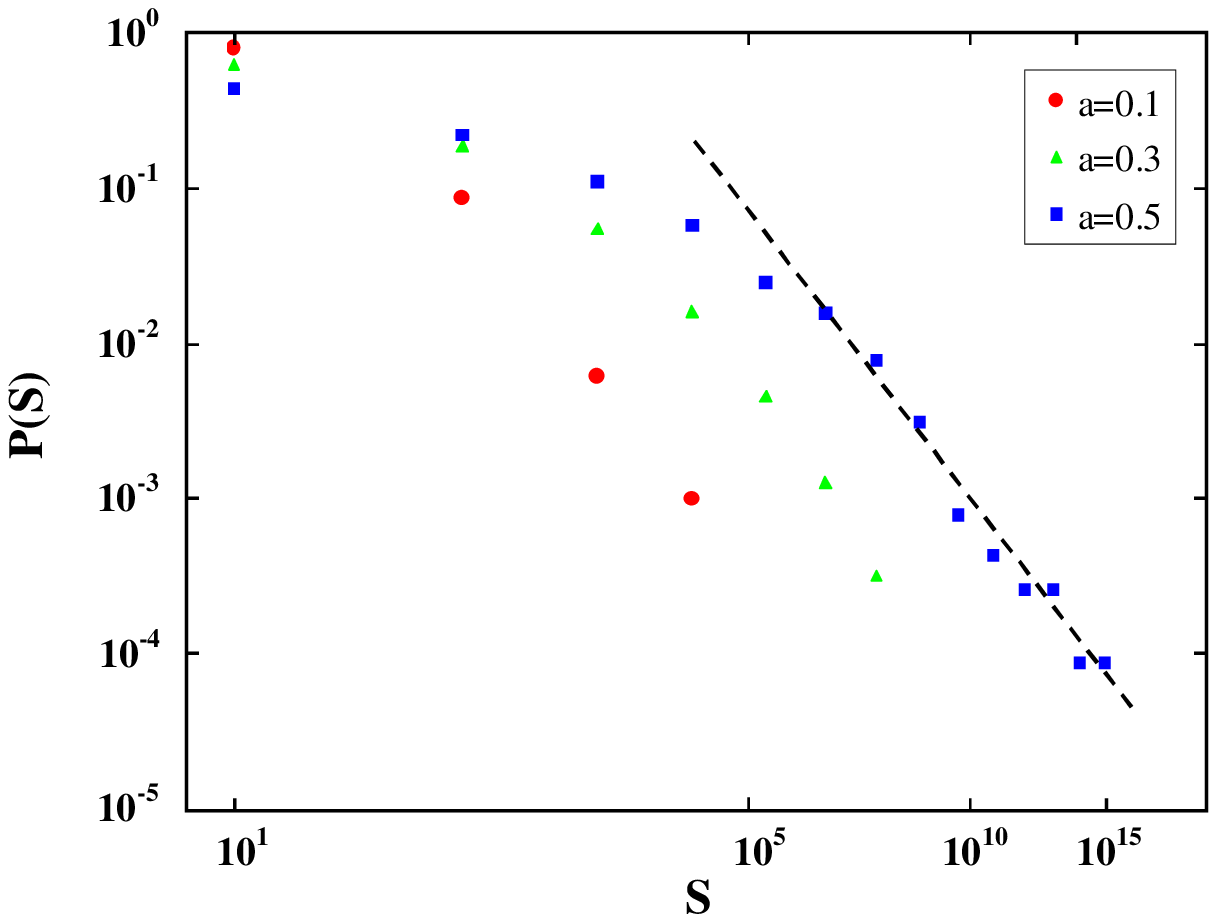}
\caption[0]{Plot of the averaged probability distribution of cluster sizes $|s|$
for the herding probability
$a=0.5$ ($h=1$), where the averaged probability distributions for the KOSPI scales as a power law
$|s|^{-\alpha}$ with the exponent $ \alpha=5.72$(the dot line).}
\end{figure}
Since the distribution of returns is directly related to the distribution of
clusters, the averaged distribution of cluster $P(s)$ scales as a power law 
\begin{equation}
P(s) \simeq |s|^{-\alpha}
\label{eq:bb12}
\end{equation}
with the scaling exponent $\alpha$. 
Fig.$3$ in our model 
present the log-log plot of the averaged
distribution of cluster against the sizes $|s|$ of the
transacted states from tick data of the KOSPI
and the scaling exponents are found to take $\alpha =7.68$(the won-dollar exchange rate)
and $5.72$(the KOSPI), remarkably different from theoretical and numerical results $[4,5]$. 
%
%
%
To find the distribution of the price return $R$ for different herding probabilities,
the herding parameter of the network of agents can be estimated from $P(\psi_l ={+1, -1})=$$a=a_{+} + a_{-} $,
where $P(\psi_l =+1)=$$a_{+}$ and $P(\psi_l =-1)=$$ a_{-}$ are, respectively, 
the probability of the selling and buying herd.
By introducing $ h \equiv \frac{1-a}{a}$ as the the herding parameter that stands for 
a measure of the herd behaviors, the active herding behavior is observed for $h>0(a<1)$,
while no herding behavior takes place for  $h=0(a=1)$
When we perform the simulation of $P(R)$, the herding parameter is incorporated into 
the price return, whose elements are the random numbers generated from the 
real data in Fig.$2$.
\begin{figure}[t]
\includegraphics[width=8.5cm]{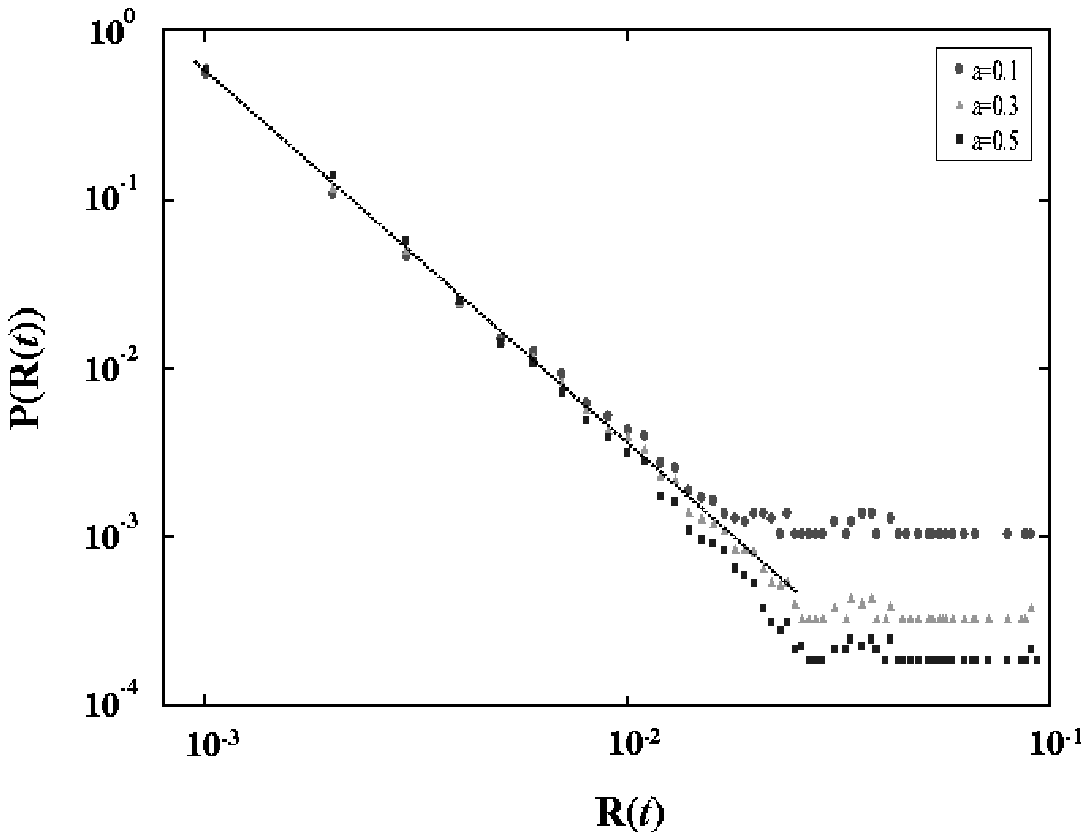}
\caption[0]{ Log-log plot of the probability distribution of returns for three types of herding probabilities
$a=0.1$, $0.3$, $0.5$ ($h=9$, $2.33$, $1$), where the dot line scales as a power law
$R^{-\beta}$ with the exponent $ \beta=2.2$ for the won-dollar exchange rate.}
\end{figure}
\begin{figure}[b]
\includegraphics[width=8.5cm]{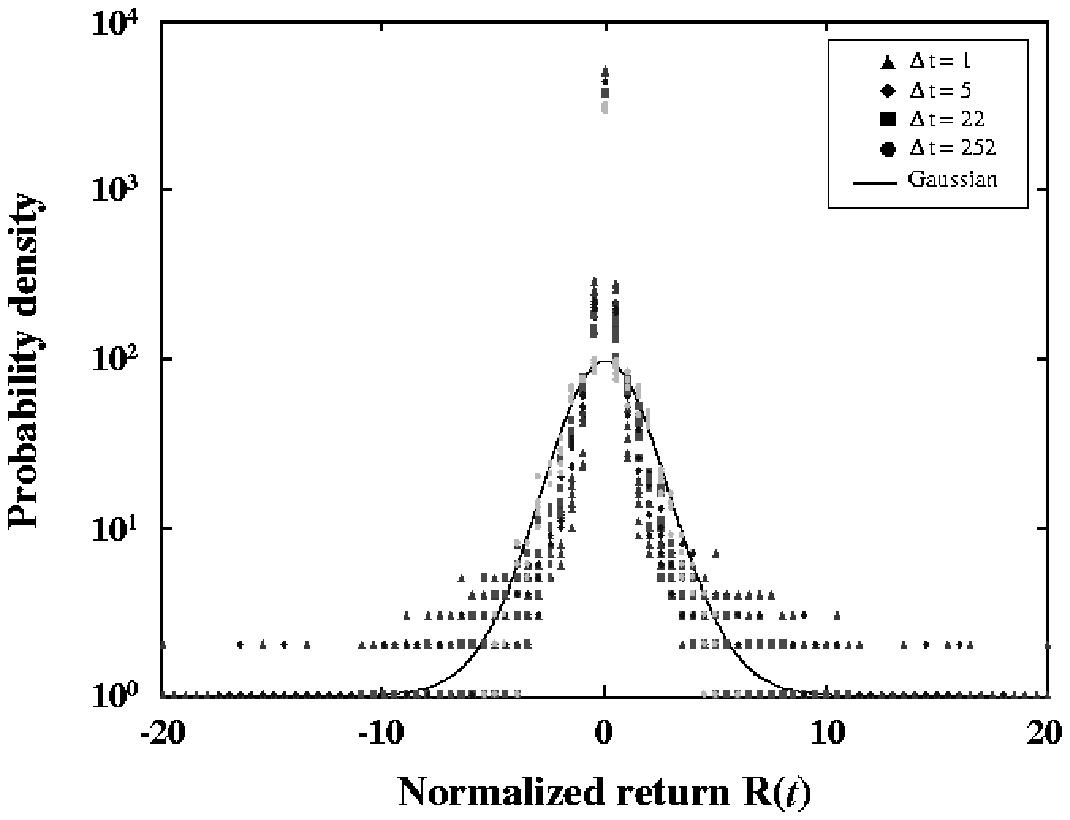}
\caption[0]{ Semi-log plot of the probability distribution of normalized returns
for herding probability $a=0.1$($h=9$),
where the dot line is the form of Gaussian function($B=2.0$ and $b =0.06$) for
the won-dollar exchange rate. }
\end{figure}

The probability distribution of returns $P(R)$ for three herding parameters
satisfies the power law
\begin{equation}
 P(R) \simeq R^{-\beta}
\label{eq:c1}
\end{equation}
with the exponents $ \beta=2.2$(the won-dollar exchange rate) and $2.4$(the KOSPI),
as shown in Fig.$4$.
Here we would suggest that $h^{*} =2.33$$(a=0.3)$ is the so-called critical herding parameter
from our data. It is obtained that the financial crashes occur at $a<0.3$($h>2.33$).
Thus the probability of extremely high returns appears to increase in the crash regime,
since the states of the transaction exists to decrease for $h>h^{*}$.
Next we calculate the distribution of normalized returns.
Since the statistical quantity $<R>$ is the value of returns
averaged over the time series of $R$
and the volatility $\sigma$ is defined as $\sigma= (<R^2 > - <R>^2 )^{1/2}$,
the normalized return $R(t)$ can be represented in terms of
\begin{equation}
  R(t) =(R-<R>)/\sigma. 
\label{eq:d1}
\end{equation}
As the time step takes the larger value, the probability distribution of normalized returns
is expected to approach to a Gaussian form, viz.
\begin{equation}
 P(R(t)) = B \exp [-b R^2 (t)]. 
\label{eq:e1}
\end{equation}
In Fig.$5$, we show
the semi-log plot of the probability distribution of the normalized returns
for the herding parameter $a=0.1$($h=9$),
where the time steps are taken as $\Delta t=1$, $5$, $22$, and $252$
for the won-dollar exchange rate.
In this case the form of the fat-tailed distribution appears 
for the time intervals $\Delta t=1$, $5$ and $22$, and the probability distribution of 
normalized returns really reduces to a Gaussian form for the time interval larger than 
$\Delta t=252$.  
Hence our results from a Gaussian form are as follows: $B=2.0$, $b =0.06$
for the won-dollar exchange rate, and $B=1.7$, $b =0.04$ for the KOSPI.

\section{Conclusions}

In conclusion, we have investigated the dynamical herding behavior 
for the won-dollar exchange rate and the KOSPI in Korean financial market.
Specially, the distribution of the price return scales as a power law
$R^{-\beta}$ with the exponents $ \beta=2.2$(the won-dollar exchange rate) and $2.4$(the KOSPI).
It is in practice obtained that our scaling exponents $ \beta $ are somewhat larger than the
numerical $1.5$ $[5]$.
It would be noted that the returns in the probability existing financial crashes
are extremely high, which the active herding behavior
occurs with the increasing probability as the herding parameter
takes larger value in real financial Markets.
We would suggest that the critical value of herding parameter$[19]$ is $h^{*} =2.33$($a=0.3$).
It is found that the distribution of normalized returns 
reduces to a Gaussian form, 
and there arises a crossover toward a Gaussian probability function
for the distribution of normalized returns.
%
%

In future, our analyses plan to investigate in detail the herd behavior
for the yen-dollar exchange rate,
and we hope that the dynamical herd behaviors apply extensively to the other tick data
in foreign financial market.

\begin{acknowledgements}
This work was supported by Grant No.R01-2000-000-00061-0 from the Basic Research 
Program of the Korea Science and Engineering Foundation.
Y. K. acknowledges the support by the Basic Research
Program of the Korea Science and Engineering Foundation
(Grant No. R01-2001-000-00025-0).
\end{acknowledgements}

\end{document}